\documentclass[aps,prl,twocolumn,superscriptaddress,showpcs]{revtex4-1}
\usepackage{times}
\usepackage{graphicx}
\usepackage{longtable}
\usepackage{color}
\usepackage{amsmath}
\usepackage{amssymb}
\usepackage{multirow}

\newcommand{\beq}{\begin{equation}}
\newcommand{\eeq}{\end{equation}}
\newcommand{\beqa}{\begin{eqnarray}}
\newcommand{\eeqa}{\end{eqnarray}}
\newcommand{\ket} [1] {\vert#1\rangle}

\def\ket#1{|#1\rangle}

\def\opone{\leavevmode\hbox{\small1\kern-3.8pt\normalsize1}}

\begin{document}

\title{Bell Experiments with Random Destination Sources}

\author{Fabio Sciarrino}
\homepage{http://quantumoptics.phys.uniroma1.it}
\affiliation{Dipartimento di Fisica della ``Sapienza''
Universit\`{a} di Roma,
Roma 00185, Italy}
\affiliation{Istituto Nazionale di Ottica (INO-CNR), L.go E. Fermi 6, I-50125 Florence, Italy}
\author{Giuseppe Vallone}
\homepage{http://quantumoptics.phys.uniroma1.it}
\affiliation{Centro Studi e Ricerche ``Enrico Fermi'', Via Panisperna 89/A, Compendio del Viminale, Roma 00184, Italy}
\affiliation{Dipartimento di Fisica della ``Sapienza''
Universit\`{a} di Roma,
Roma 00185, Italy}
\author{Ad\'an Cabello}
\affiliation{Departamento de F\'{\i}sica
Aplicada II, Universidad de Sevilla, E-41012 Sevilla, Spain}
\author{Paolo Mataloni}
\homepage{http://quantumoptics.phys.uniroma1.it}
\affiliation{Dipartimento di Fisica della ``Sapienza''
Universit\`{a} di Roma,
Roma 00185, Italy}
\affiliation{Istituto Nazionale di Ottica (INO-CNR), L.go E. Fermi 6, I-50125 Florence, Italy}

\date{\today}


\begin{abstract}
It is generally assumed that sources sending randomly two
particles to one or two different observers, named here
{random destination sources (RDS)}, cannot by used
for genuine quantum nonlocality tests because of the
postselection loophole. We demonstrate that Bell experiments
not affected by the postselection loophole may be performed
with: (i) RDS and local postselection using perfect detectors,
(ii) RDS, local postselection, and fair sampling assumption
with any detection efficiency, and (iii) RDS and a threshold
detection efficiency required to avoid the detection loophole.
These results allow the adoption of RDS setups, which are more
simple and efficient, for long-distance free-space Bell tests,
{and extends the range of physical systems which can
be used for loophole-free Bell tests.}
\end{abstract}


\pacs{03.65.Ud, 42.65.Lm, 03.67.Bg, 42.50.Xa}

\maketitle


{\it Introduction.---}An experimental loophole-free violation
of a Bell inequality is of fundamental importance not only for
ruling out the possibility of describing nature with local
hidden variable theories \cite{bell64phy}, but also for proving
entanglement-assisted reduction of classical communication
complexity \cite{buhr10rmp}, device-independent eternally
secure communication \cite{acin07prl}, and random number
generation with randomness certified by fundamental physical
principles \cite{piro10nat}. {This explains the interest in
loophole-free Bell test over long distances.} There are three
types of loopholes. The locality loophole \cite{bell81jpc}
occurs when the distance between the local measurements is too
small to prevent causal influences between one observer's
measurement choice and the other observer's result. To avoid
this possibility these two events must be spacelike separated.
The detection loophole \cite{pear70prd} occurs when the overall
detection efficiency is below a minimum value, so although the
events in which both observer's have results might violate the
Bell inequality, there is still the possibility to make a local
hidden variable model which reproduces all the experimental
results. To avoid this possibility the overall detection
efficiency must be larger than a threshold value. Finally, the
postselection loophole \cite{aert99prl, cabe09prl, deca94pra}
occurs when the setup does not always prepare the desired
state, so the experimenter postselects those events with the
required properties. It has been shown that, in certain
configurations, the rejection of ``undesired'' events can be
exploited by a {{\em local}} model to imitate the predictions
of quantum mechanics \cite{aert99prl}. However, it has been
recently pointed out that the postselection loophole is not due
to the rejection of undesired events itself, but rather to the
geometry of the setup. The loophole can be fixed with a
suitable geometry, without renouncing to the postselection
\cite{cabe09prl, lima10pra}. This leads to the question of when
Bell experiments with postselection are legitimate. The answer
is interesting since sources with postselection can be simpler
and more efficient.

In this paper we analyze schemes which are believed to be
affected by the postselection loophole \cite{deca94pra,
pope97pra,kwia94pra}. In particular, we focus on random
destination sources (RDS) emitting two photons, such that
sometimes one photon ends in Alice's detectors and the other in
Bob's, but sometimes both end in the same party's detectors.
{The same argument can be used with 
massive particles, as discussed later}.
Hereafter we will consider as benchmark the source shown in
Fig.~\ref{fig:source} \cite{mart09ope}: two photons with
horizontal and vertical polarization are generated via
collinear spontaneous parametric down conversion (SPDC) and a
beamsplitter (BS) splits the photons over the modes $A$ and
$B$. Three different cases may occur: Both photons emerge on
mode $A$ ($\ket{HV}_A$), both on mode $B$ ($\ket{HV}_B$), or
the two photons are divided into different modes
[$(\ket{H}_A\ket{V}_B-\ket{V}_A\ket{H}_B)/\sqrt{2}$]. {Such a
setup is a natural candidate for long-distance free-space
experiments requiring quantum entanglement, and specifically
for satellite-based quantum communications \cite{ursi08iac,
jin10npho}, since it satisfies the requirements of high
efficiency (due to the adoption of periodically poled
crystals), stability, compactness (the beam splitter could even
be manufactured onto a single chip with the SPDC source), and
emission over a single spatial mode.

\begin{figure}[t]
 \centering
 \includegraphics[width=8cm]{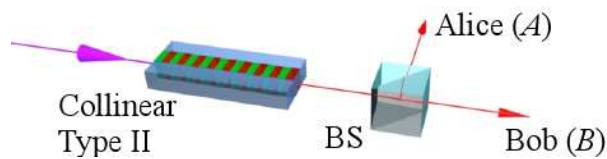}
 \caption{Collinear source of two photon states. The two particles
 are produced by type II phase matching and randomly broadcasted to two observers $A$ and $B$
 via a beam splitter (BS).}
 \label{fig:source}
\end{figure}


We will show that Bell experiments not affected by
postselection loophole can be performed in these cases: (i) RDS
and local postselection with perfect detectors. (ii) RDS, local
postselection and fair sampling assumption for any value of
detection efficiency. (iii) RDS and a threshold detection
efficiency required to avoid even the detection loophole.


Let us consider an experiment with RDS producing two photons in
different locations, Alice (A) and Bob (B), with probability
$p$, two photons in Alice's side with probability
$\frac{1-p}{2}$, and two photons on Bob's side with probability
$\frac{1-p}2$. Detection efficiencies in Alice and Bob's sides
are $\eta_A$ and $\eta_B$, respectively. Alice and Bob also
have photon number discriminators. The probability $p$ will
depend on the particular configuration. For instance, in the
case of the source of Fig.~\ref{fig:source}, $p=2T(1-T)$, where
$T$ is the transmittance of the beam splitter.

{\it (i) Perfect detectors and local postselection.---}Let's
consider $\eta_A=\eta_B=1$. Alice and Bob, depending on the
measured number of photons, locally decide to postselect only
those events in which entanglement has been successfully
distributed. See column $\eta=1$ of Table~\ref{tableI}. Is such
local discarding of events introducing any loophole?

The answer is no, since the selection and rejection of events is
independent of the local measurement settings (otherwise a
local hidden variables model could exploit the selection to
violate the inequality).
Indeed, any Bell experiment with local postselection, in the sense that
it does not require Alice and Bob to communicate, is free of
the postselection loophole. Local postselection is not a
\textit{necessary requirement} to be free of the postselection
loophole (see \cite{aert99prl} for a counterexample), but is a
\textit{sufficient property} to be free of this loophole.


\begin{figure}[t]
 \centering
 \includegraphics[width=7cm]{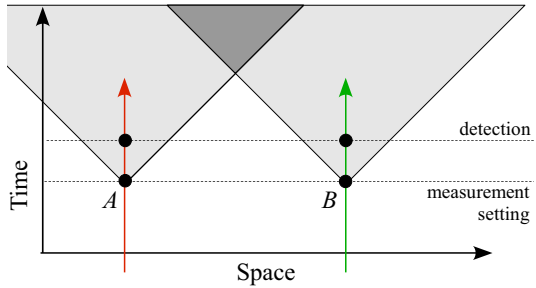}
 \caption{Space-time diagram for loophole-free Bell experiments.}
 \label{fig:diagram}
\end{figure}


This is a crucial point which deserves a detailed examination.
First consider a {\em selected} event: The two photons have
been detected at different locations, corresponding one to
Alice's detector $D_A$ and the other to the Bob's detector
$D_B$. If the detection in $B$ is outside the forward
light-cone of the measurement setting in $A$ (this is precisely
the locality condition), no mechanism could turn a rejected
event into a selected one (see Fig.~\ref{fig:diagram}). Let us
consider a {\em rejected} event, for example when two photons
have been detected at $D_A$. Both Alice (since she registers a
double detection) and Bob (since he does not register any
detection) locally discard the event. Again, due to the
locality condition, the double detection at Alice's side cannot
be caused by Bob's measurement setting, and the absence of Bob
detection cannot be influenced by Alice's measurement setting.
The same happens when two photons go towards Bob's side.

Hence, when Alice or Bob locally discard the events, there is
no physical mechanism preserving locality which can turn a
selected (rejected) event into a rejected (selected) event. The
selecting of events is {\em independent} of the local settings.
For the selected events only the result can depend on the local
settings. This is exactly the condition under which the Bell's
inequalities are valid. Therefore, an experimental violation of
them based on local postselection with a random destination
source provides a conclusive test of local realism when perfect
detectors are used.


\begin{table}[t]
\begin{ruledtabular}
\begin{tabular}{c|cc||c|ccccc}
\multicolumn{3}{c||}{GENERATION}&\multicolumn{6}{c}{DETECTION}\\
\hline
\multicolumn{3}{c||}{}&$\eta=1$&\multicolumn{5}{c}{$\eta\neq1$}\\
&$A$ & $B$ &CHSH& $A$ & $B$ &CHSH & CH &CHSH\\
&&&(no-fs)&&&(fs) & (no-fs) &(no-fs)\\
\hline
\multirow{3}{*}{Events I} &\multirow{3}{*}{1} & \multirow{3}{*}{1} &\multirow{3}{*}{\checkmark}&
1 & 1 & \checkmark& \checkmark& \checkmark\\
&&&&1 & 0 & NO& \checkmark& \checkmark\\
&&&&0 & 1 & NO& \checkmark& \checkmark\\
&&&&0 & 0 & NO& $\times$& \checkmark\\
\hline
\multirow{2}{*}{Events II} &\multirow{2}{*}{2} & \multirow{2}{*}{0} &\multirow{2}{*}{NO}&
2 & 0 & NO& $\times$& \checkmark\\
&&&&1 & 0 & NO& \checkmark& \checkmark\\
&&&&0 & 0 & NO& $\times$& \checkmark\\
\hline
\multirow{2}{*}{Events III} &\multirow{2}{*}{0} & \multirow{2}{*}{2} &\multirow{2}{*}{NO}&
0 & 2 & NO& $\times$& \checkmark\\
&&&&0 & 1 & NO& \checkmark& \checkmark\\
&&&&0 & 0 & NO& $\times$& \checkmark\\
\end{tabular}
\caption{Different types of events (I, II and III), depending
on the number of photons sent to observers $A$ and $B$.
Different nonlocality tests are considered according to the
detection efficiency ($\eta$), the adopted inequality (CH,
CHSH) and whether fair-fampling (fs) is assumed. The numbers
refers to the photons generated or detected by each observer.
NO: Discarded events, $\times$: Events not discarded but not
contributing to the inequality, $\checkmark$: Events which
contribute to the inequality.}
 \label{tableI}
\end{ruledtabular}
\end{table}



({\it ii) Fair sampling assumption and local
postselection.---}Let's now consider to case of imperfect
detection efficiency of the measurement apparatus. Some of the
events are lost, and Alice and Bob only keep coincidences and
assume fair sampling (\emph{i.e.}, that the coincidences are a
statistically fair sample of the pairs in which one photon has
gone to Alice and the other to Bob). Under this assumption,
Bell experiments based on postselection are able to show
violations of Bell's inequalities. Indeed, in the case of
$p=1$, the fair sampling assumption allows Alice and Bob to
discard the contribution where just one particle is detected.
When $p\neq1$, we have already shown that the contribution due
to double particle detection on the same observer can also be
discarded. Hence, when fair sampling is assumed, there is no
difference between Bell experiments with or without local
postselection.


{\it (iii) Threshold detector efficiency for loophole-free test
with RDS.---}What happens when the fair sampling assumption is
not considered? We will show that by considering all the
possible events, \emph{i.e.}, without postselection or fair
sampling, a loophole-free violation with RDS can be obtained
with a suitable threshold detection efficiency. For the
Clauser-Horne (CH) \cite{clau74prd} and the
Clauser-Horne-Shimony-Holt (CHSH) Bell inequalities
\cite{clau69prl}, this threshold can be obtained as follows.

In the perfect detection scenario (i), all the events in which
two particles are sent to the same observer (events II and III
of Table~\ref{tableI}) are discarded. Here, due to
inefficiency, some of the events II and III will contribute to
the data and cannot be locally discarded (see the second column
of Table~\ref{tableI}).


\begin{figure}[t]
 \centering
 \includegraphics[width=7cm]{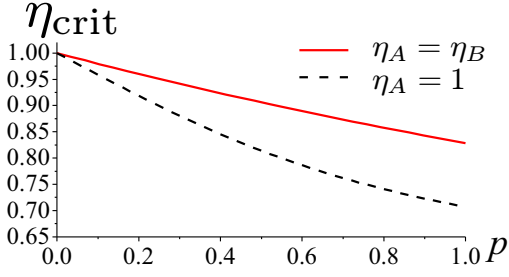}
 \caption{Critical efficiency for a loophole-free Bell test in the symmetric ($\eta_A\,$=$\,\eta_B$$>$$\eta_{\rm crit}$)
 or in the asymmetric ($\eta_A\,$=$\,1$, $\eta_B$$>$$\eta_{\rm crit}$) case.}
 \label{fig:CHSHeta}
\end{figure}


Let us consider two observers, Alice and Bob, with dichotomic
observables $a_i=\pm1$ and $b_i=\pm1$, respectively.
Any theory
assuming realism and locality must satisfy the CH inequality,
\begin{equation}
\begin{aligned}
 I_{\rm CH}=&p(a_1,b_1)+p(a_2,b_1)+p(a_1,b_2)\\
 &-p(a_2,b_2)-p(a_1)-p(b_1)\leq0,
 \end{aligned}
\label{CH}
\end{equation}
where $p(a_i,b_j)$ is the probability that Alice obtains
$a_i=1$ and Bob obtains $b_j=1$, while $p(a_1)$ is the
probability that Alice obtains $a_1=1$. We adopt one detector
in each side corresponding to the $+1$ outcome. We set $a_i=+1$
($b_i=+1$) when Alice (Bob) detects only one photon, while
$a_i=-1$ ($b_i=-1$) when Alice (Bob) detects zero or two
photons. In this way, the inequality is insensitive to any
normalization implying that the ``vacuum contribution'' of
standard SPDC sources does not contribute to $I_{\rm CH}$.
Moreover, the events in which two particles are detected by
Alice and no particle by Bob (or viceversa), indicated by
$\times$ in Table \ref{tableI}, do not contribute to Eq.
\eqref{CH}. Indeed \eqref{CH} involves only detection events in
which at least one observable is +1.

Let us define $Q$ as the value of $I$ corresponding to the case
when both particles are detected, $M_A$ ($M_B$) the value of
$I$ when only particle $A$ ($B$) is detected from the (1 1)
events, $T_A$ ($T_B$) the value of $I$ when only particle $A$
($B$) is detected from (2 0) and (0 2) events, $D_A$ ($D_B$)
the value of $I$ when two particles are detected at side $A$
($B$), and $X$ the value when no particle is detected. Then,
the average value of $I$ will be
\begin{equation}
 \label{general_ineq}
\begin{aligned}
 \langle I\rangle=&\eta^2_A\frac{1-p}{2}(D_A-2T_A+X)+\eta_A[pM_A+(1-p)T_A-X]\\
 &\eta^2_B\frac{1-p}{2}(D_B-2T_B+X)+\eta_B[pM_B+(1-p)T_B-X]\\
 &+\eta_A\eta_B\,p(Q-M_A-M_B+X)+X\,.
\end{aligned}
\end{equation}
It is easy to show that, for the singlet entangled state and
choosing the observables $a_i$ and $b_i$ that maximally violate
the inequality, we obtain the following values for the CH
inequality: $Q=\frac{1}{\sqrt2}-\frac 12$, $M_A=M_B=-\frac 12$,
and $X=0$. When the two particles are detected by Alice (Bob),
we have $a_i=-1$ ($b_i=-1$), which implies $D_A=D_B=0$. In
order to calculate $T_A$ ($T_B$), it is necessary to know the
particular two-photon state sent to Alice (Bob). In most RDS,
when two photons are sent to the same observer they have
orthogonal polarizations. This implies that when only one
photon is detected, we have $T_A=T_B=-\frac 12$. The local
realistic bound is violated when $\langle I\rangle
>0$:
\begin{gather}
\frac{1-p}{2}(\eta^2_A+\eta^2_B)+p\;\eta_A\eta_B(\frac12+\frac{1}{\sqrt2})-\frac12(\eta_A+\eta_B)>0.
\label{CH_condition}
\end{gather}
Note that the (0 0), (2 0), and (0 2) events do not contribute
to any term in \eqref{CH}.


In the symmetric case ($\eta_A$=$\eta_B$=$\eta$), the minimum
detection efficiency is
\begin{equation}
 \eta>\eta_{\rm crit}\equiv\frac{2}{2+p(\sqrt2-1)}.
\end{equation}
For $p=1$, we recover $\eta>\frac{2}{1+\sqrt2}\simeq 0.83$
\cite{garg87prd}. For $p=0.5$, we obtain $\eta > 0.90$. RDS
imposes a stricter constraint on the experimental setting, but
still a loophole-free nonlocality test can be achieved.


\begin{figure}[t]
 \centering
 \includegraphics[width=6cm]{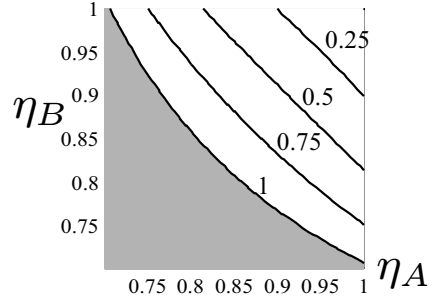}
 \caption{Allowed values of $\eta_A$ and $\eta_B$ for a loophole-free Bell test for different values of $p$
 (1, 0.75, 0.5, 0.25). For each value of $p$ the allowed zone is in the upper-right part of the
 corresponding curve.}
 \label{fig:graph}
\end{figure}


For the fully asymmetric case ($\eta_A=1$), we have $\eta_B
> \eta_{\rm crit}$ with
\begin{equation}
\label{CH_asymm}
 \eta_{\rm crit}=\frac{-1+p(1+\sqrt2)-\sqrt{4p(1-p)+(1-p-\sqrt2p)^2}}{2(p-1)}.
\end{equation}
In the limit $p\rightarrow1$ we recover
$\eta_B>\frac{1}{\sqrt{2}}\simeq0.71$ \cite{cabe07prl1}. In
Fig.~\ref{fig:CHSHeta} we show the critical values of the
efficiency in the symmetric and totally asymmetric cases. For
the general case \eqref{CH_condition}, Fig.~\ref{fig:graph}
shows the values of $\eta_A$ and $\eta_B$ allowing a
loophole-free Bell test for different values of $p$.

It is worth noting that a completely equivalent result is
obtained by using the CHSH inequality,
\begin{equation}
 \label{CHSH}
 I_{CHSH}=\langle a_1b_1\rangle+\langle a_2b_1\rangle+\langle a_1b_2\rangle-\langle a_2b_2\rangle-2\leq0,
\end{equation}
by using the arguments given in \cite{pope97pra}. When one
observer detects no particle or two particles, he sets $a_i$
($b_i$) $= +1$. When Alice detects two photons and Bob no
photon [the (2 0) events], we have $\langle
a_1b_1\rangle=\langle a_2b_1\rangle=\langle
a_1b_2\rangle=\langle a_2b_2\rangle=1$ [and similarly for the
(0 2) events]. The same happens for the (0 0) events (where
neither Alice nor Bob detects a particle). If the source
produces the singlet state when the two particles are sent to
different observers, inequality \eqref{CH_condition} holds.


\begin{figure}[t]
 \centering	
 \includegraphics[width=8cm]{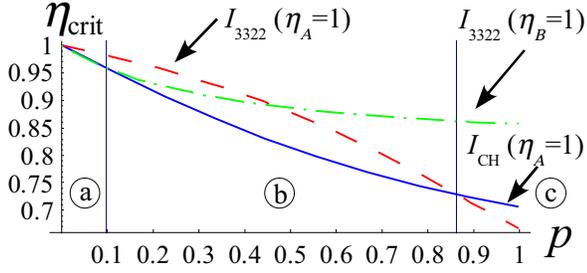}
 \caption{Critical detection efficiencies in the completely asymmetric case for the CH and the Collins-Gisin inequalities.
 $I_{3322}$ with $\eta_A=1$ (dashed line) and with $\eta_B=1$ (dash-dotted line), CH (Continuous line).
 Labels (a), (b), and (c) identify the three ranges where different inequalities lead to the lower efficiency threshold.}
 \label{fig:ch_gisi}
\end{figure}


Finally, we demonstrate that, in the asymmetric case
$\eta_A\neq\eta_B$, an inequality with lower bound with respect
to \eqref{CH_asymm} does exist for some values of $p$. Consider
the $I_{3322}$ inequality \cite{coll04jpa},
\begin{equation}
 \label{gisin}
\begin{aligned}
 I_{3322}=&p(a_1,b_1)+p(a_1,b_2)+p(a_1,b_3)+p(a_2,b_1)\\
 &+p(a_2,b_2)+p(a_3,b_1)-p(a_2,b_3)-p(a_3,b_2)\\
 &-2p(a_1)-p(a_2)-p(b_1)\leq0.
\end{aligned}
\end{equation}
By setting $a_i=1$ ($b_i=1$) when Alice (Bob) detects zero or
two photons, it is possible to show that the inequality is
violated if
\begin{equation}
 \label{gisin2}
 \frac{1-p}{2}(\eta^2_A+3\eta^2_B)+p\;\eta_A\eta_B\frac94-\frac12(\eta_A+3\eta_B)>0.
\end{equation}
Eq. \eqref{gisin2} depends in a different way on $\eta_A$ and
$\eta_B$, hence we will consider separately the two conditions
$\eta_A=1$ and $\eta_B=1$. We may compare the efficiency
threshold in Fig.~\ref{fig:ch_gisi}. The plot is divided in
three regions (a--c), depending on which inequality leads to
the lowest efficiency threshold.

For $\eta_A=1$, the lower bound on $\eta_B$ is
$\eta_B>4p/(9p-6+\sqrt{36-60p+33p^2})$, which is better than
\eqref{CH_asymm} for any $p>0.863$ (c). For $p=1$, we obtain
the same results presented in \cite{brun07prl}. By setting
$a_i=-1$ ($b_i=-1$), when Alice (Bob) detects zero or two
photons, we obtain the same result with
$\eta_A\leftrightarrow\eta_B$. In this case, for $\eta_A=1$,
the lower bound on $\eta_B$ is better than the CH condition for
any $p<0.099$ (see Fig.~\ref{fig:ch_gisi}) (a). In the central
region (b), CH is still the optimal choice. Due to the $T_A$
and $T_B$ terms in \eqref{general_ineq}, the efficiency bound
depends on the specific form of the source. Here we have
calculated the bound for the case of two photons sent to the
same observers with orthogonal polarization.

{Finally, RDS are useful for loophole-free Bell
tests beyond the most promising proposals to date (see
\cite{ros09asl} and references therein). The idea is to combine
RDS of pairs of massive particles such as neutrons or molecules
with interferometric setups like \cite{cabe09prl} into Bell
experiments with postselection, and take advantage of the fact
that the detection efficiencies for these particles are above
the thresholds obtained in this paper. So far, the scheme in
\cite{cabe09prl} has been tested with photons \cite{lima10pra}
and electronic currents \cite{frus09prb}, but there seems to be
no fundamental problem in performing similar experiments with
molecules (see \cite{gnei08prl} for an example of a RDS with
molecules) and accelerator-based sources of neutrons
\cite{has10com}.}


This work was supported by FIRB Futuro in Ricerca - HYTEQ. A.C.
acknowledges support from the Spanish MCI Project No.\
FIS2008-05596.


\begin{thebibliography}{10}
\providecommand{\url}[1]{\texttt{#1}}
\providecommand{\urlprefix}{URL }
\providecommand{\eprint}[2][]{\url{#2}}

\bibitem{bell64phy}
J.~S. Bell, Physics \textbf{1}, 195 (1964).

\bibitem{buhr10rmp}
H.~Buhrman, R.~Cleve, S.~Massar, and R.~de~Wolf, Rev. Mod. Phys. \textbf{82},
  665 (2010).

\bibitem{acin07prl}
A.~Ac\'\i{}n \emph{et~al.}, Phys. Rev.
    Lett. \textbf{98}, 230501 (2007).

\bibitem{piro10nat}
S.~Pironio \emph{et~al.}, Nature (London)
    \textbf{464}, 1021 (2010).

\bibitem{bell81jpc}
J.~Bell, J. Phys. C \textbf{2}, 41 (1981).

\bibitem{pear70prd}
P.~M. Pearle, Phys. Rev.~D \textbf{2}, 1418 (1970).

\bibitem{aert99prl}
S.~Aerts, P.~Kwiat, J.-\AA. Larsson, and
    M.~\.{Z}ukowski, Phys. Rev. Lett.
  \textbf{83}, 2872 (1999).

\bibitem{cabe09prl}
A.~Cabello \emph{et~al.}, Phys. Rev. Lett.
    \textbf{102}, 040401 (2009).

\bibitem{deca94pra}
L.~{De Caro} and A.~Garuccio, Phys. Rev. A \textbf{50}, R2803 (1994).

\bibitem{lima10pra} G.~Lima \emph{et~al.}, Phys. Rev. A
    \textbf{81}, 040101(R) (2010).

\bibitem{pope97pra}
S.~Popescu, L.~Hardy, and M.~\ifmmode~\dot{Z}\else \.{Z}\fi{}ukowski, Phys.
  Rev. A \textbf{56}, R4353 (1997).

\bibitem{kwia94pra}
P.~G. Kwiat, P.~H. Eberhard, A.~M. Steinberg, and R.~Y. Chiao, Phys. Rev. A
  \textbf{49}, 3209 (1994).

\bibitem{mart09ope}
A.~Martin \emph{et~al.}, Opt. Express
    \textbf{17}, 1033 (2009).

\bibitem{ursi08iac}
R.~{Ursin \it et al.}, IAC Proceedings \textbf{A2.1.3} (2008),
  \eprint{quant-ph/0806.0945}.

\bibitem{jin10npho}
X.-M. Jin \emph{et~al.}, Nat. Photon.
    \textbf{4}, 376 (2010).

\bibitem{clau74prd}
J.~F. Clauser and M.~A. Horne, Phys. Rev. D \textbf{10}, 526 (1974).

\bibitem{clau69prl}
J.~F. Clauser, M.~A. Horne, A.~Shimony, and R.~A. Holt, Phys. Rev. Lett.
  \textbf{23}, 880 (1969).

\bibitem{garg87prd}
A.~Garg and N.~D. Mermin, Phys. Rev. D \textbf{35}, 3831 (1987).

\bibitem{cabe07prl1}
A.~Cabello and J.-\AA. Larsson, Phys. Rev.
    Lett. \textbf{98}, 220402 (2007).

\bibitem{coll04jpa}
D.~Collins and N.~Gisin, J. Phys. A
  \textbf{37}, 1775 (2004).

\bibitem{brun07prl}
N.~Brunner, N.~Gisin, V.~Scarani, and C.~Simon, Phys. Rev. Lett. \textbf{98},
  220403 (2007).

\bibitem{ros09asl}
 W. Rosenfeld {\it et al.},
 Adv. Sci. Lett. \textbf{2}, 469 (2009).

\bibitem{frus09prb}
 D. Frustaglia and A. Cabello,
 Phys. Rev. B \textbf{80}, 201312(R) (2009).

\bibitem{gnei08prl}
 C. Gneiting and K. Hornberger,
 Phys. Rev. Lett. \textbf{101}, 260503 (2008).

\bibitem{has10com}
 Y. Hasegawa (private communication).



\end{thebibliography}

\end{document}